\documentclass[aps,twocolumn,showpacs,superscriptaddress]{revtex4}
\usepackage{amsbsy}
\usepackage{bm}
\usepackage{graphicx}

\newcommand{\nc}{\newcommand}
\nc{\da}{\dagger}
\nc{\noi}{\noindent}     
\nc{\eq}[1]{\mbox{Eq.(\ref{#1})}}
\nc{\ba}{\begin{array}}
\nc{\ea}{\end{array}}
\nc{\bea}{\begin{eqnarray}}
\nc{\eea}{\end{eqnarray}}
\nc{\fig}[1]{\mbox{Fig.~\ref{#1}}}

\nc{\bc}{\begin{center}}
\nc{\ec}{\end{center}}

\begin{document}

\title{On the  Theory of Casimir-Polder Forces}

\author{Bo-Sture Skagerstam}\email{bo-sture.skagerstam@ntnu.no}
\author{Per Kristian Rekdal}\email{per.rekdal@ntnu.no}
\author{Asle Heide Vaskinn}\email{asle.vaskinn@ntnu.no}
\affiliation{Department of Physics, 
             The Norwegian University of Science and Technology, N-7491 Trondheim, Norway}


\begin{abstract}

     We consider the energy shift for an atom close to a non-magnetic body with a magnetic moment coupled to a quantized magnetic field. The corresponding {\it repulsive} Casimir-Polder force is obtained for a perfect conductor, a metal, a dielectric medium, with dielectric properties modeled by a Drude formula,  and a superconductor at zero temperature. The dielectric properties of the superconductor is obtained by making use of the Mattis-Bardeen linear response theory and we present some useful expressions for the low-frequency conductivity.  The quantum dynamics with  a given initial state is discussed in terms of the well-known Weisskopf-Wigner theory and is  compared with corresponding results for a electric dipole coupling. The results obtained are compatible with a conventional master equation approach. In order to illustrate the dependence on geometry and material properties, numerical results are presented for the ground state using a two-level approximation.

\end{abstract}
\maketitle


\bc{
\section{INTRODUCTION}
\label{sec:introd}}
\ec


      In recent years \cite{parsegian_05}, great advances in experimental techniques have stimulated an intense theoretical as well as experimental activity on the Casimir effect \cite{casimir_48}. In particular, the study of the force between an atom and a bulk material - the Casimir-Polder (CP) force \cite{casimir_polder_48,lifshitz_61} - at finite temperature has recently become a popular subject of research (see e.g. Refs.\cite{Klimchitskaya_08,Lamoreaux_07,buhmann_08, Lamoreaux_08} and references therein). The CP force on e.g. a Bose-Einstein condensate is, in fact, not only   of theoretical  interest as the thermal corrections are measurable \cite{haroche_2008,obrecht_07}. CP forces play an important role in a variety of processes in physical chemistry, atom optics, and cavity QED. This includes e.g. nano-technological applications \cite{Chan_01,klimov_09}, the possibility of investigating fundamental forces at the sub-micrometer scale \cite{Dimopoulos_03} and the effort to miniaturize atom chips \cite{folman_07,cano_2008}. The CP force can, under some circumstances, turn from an attractive character to a repulsive one leading to a quantum levitation phenomena. This can be achieved using left-handed meta-materials \cite{leonhardt_2007} or if the system under consideration is immersed in a suitable fluid \cite{munday_2009}. As we will argue in the present paper one can easily obtain repulsive CP forces if one consider magnetic moment transitions instead of electric dipole ones without the need of additional ingredients. Such repulsive forces tend, however, to be very small as compared to the conventional attractive electric dipole CP forces.

\begin{figure}[t]

\begin{picture}(0,0)(175,450)   
\includegraphics{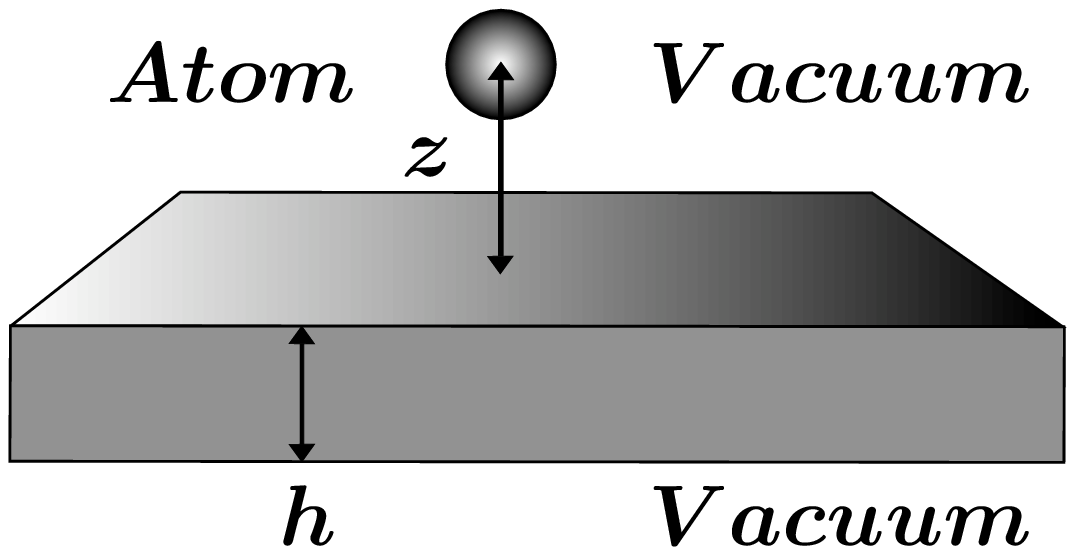}
\end{picture}
\vspace{5cm}
\caption{Schematic picture of the setup considered in our calculations. An atom inside a microtrap is 
         located in vacuum at a distance $z$ away from a  non-magnetic slab with thickness $h$.
	 The slab has an infinite extension in the other directions.
         Vacuum is on both sides of the slab. The  slab can e.g. be a normal conducting metal or a
         superconducting metal.}
\label{geo_slab_fig}
\end{figure}

Experiments reveal that the proximity of atoms to the non-magnetic body may introduce several surface-related decoherence effects (see e.g. Refs.\cite{hinds_03,harber_03,vuletic_04}). Most importantly, atoms may be expelled from an atom trap due to spin flip transitions induced by Johnson-noise currents in the material. If the expelled atoms can not re-enter the trap, it is not clear to what extent the system in general can be described in terms of a conventional master equation. 

Specifically we will consider a multilevel atom close to a non-magnetic slab as illustrated in Fig.\ref{geo_slab_fig}, where the atom may be held close to the slab by e.g. a microscopic atom trap \cite{folman_07}.  We focus on magnetic moment couplings to the electromagnetic field but we will also mention, where appropriate, relevant effects for couplings to electric dipole moments.  Furthermore, applying the Heisenberg operator equations of motion and the rotating-wave approximation lead in general to operator ordering problems and great care has to be taken in order to avoid mathematical inconsistencies (see e.g. Ref.\cite{barnett_97}). The well-known Weisskopf-Wigner theory \cite{wigner_30} will, however, describe the physics in a more clear manner and, as we will see,  clarifies the role of various approximative procedures. 
Numerical results will be presented by making use of a two-level approximation. The two-level approximation for electric dipole transitions must be handled with some care due to the existence of sum rules \cite{barton_74}.  For magnetic transitions we will, however, not encounter such issues. 

 The paper is organized as follows. In the next section we outline the theoretical framework. In order to obtain the energy shift for ground state atoms it is argued that one has to go beyond the  conventional rotating-wave approximation. In Section \ref{sec_half} we give explicit expressions for the appropriate Green's functions in the case of a semi-infinite slab and magnetic transitions. Explicit expressions for the ground state energy shift are then obtained for a zero temperature  perfect conductor,  a metallic slab with dielectric properties described by a Drude dispersion relation and extensions of it, a dielectric medium, and for a superconductor as described by a weakly coupled BCS superconductor. In the case of a superconducting slab we have to reconsider the low-frequency dielectric properties in great detail including the presence of impurities. The short and long distance behavior of the corresponding CP force, as induced by a magnetic moment, are considered in detail. Numerical results  are compared with the corresponding results for electric dipole transitions. By defining a suitable rescaled and dimensionless CP force, we can easily compare some of our results with similar results for CP forces as induced by electric dipole transitions.  In Section  \ref{sec:summary} we give some final remarks concerning electric dipole transitions. Technical details concerning the Green's functions used for magnetic transitions, their analytical continuation as well as  various explicit expansion for a perfect conductor are presented in two appendices.

\vspace{1.0cm}
\bc{
\section{GENERAL THEORY}
\label{sec_gen}}
\ec
\vspace{-0.5cm}

   Let us consider a neutral atom at a fixed position ${\bf r}_A$.
   The magnetic moment of the atom interacts with the quantized magnetic field via a conventional Zeeman coupling.
   The total Hamiltonian is then
\bea   \nonumber
      &&~~~~~~~~~~~~ H  =      \sum_{\alpha}  E_{\alpha} \, | \alpha \rangle \langle \alpha | 
                   \\  \label{H}
		   &+& 
		   \int d^3r\int_{0}^{\infty} d\omega \,  \hbar \omega \, \hat{\bf f}^{\da}({\bf r},\omega) \cdot
                                                                              \hat{\bf f}({\bf r},\omega) ~ + ~ H' ~,
\eea
   \noi
   where the effective interaction part is
\bea   \label{H_prime}
  H'  &=&    -   \sum_{\alpha } \sum_{\beta} \, |\alpha \rangle \langle \beta| \; {\bm{\mu}}_{\alpha \beta}  \cdot  {\bf B}({\bf r}_A) 
		    ~~ .
\eea
  \noi
  Here $\hat{\bf f}({\bf r},\omega)$ is an annihilation operator for the quantized magnetic field,
  $|\alpha \rangle$ denotes the atomic state and $E_{\alpha}$ is the corresponding energy.
     We assume non-degenerate states, i.e. $E_{\alpha} \neq E_{\beta}$ for $\alpha \neq \beta$.
   The magnetic moment of the atom is  ${\bm{\mu}}_{\alpha \beta} =  \langle \alpha | \hat{\bm{\mu}} | \beta \rangle$, where 
   $ \hat{\bm{\mu}}$ is the magnetic moment operator.
   The magnetic field ${\bf B}({\bf r}) = {\bf B}^{(+)}({\bf r}) +  {\bf B}^{(-)}({\bf r})$
   is written ${\bf B}^{(+)}({\bf r}) =  \nabla \times {\bf A}^{(+)}({\bf r})$, where 
   ${\bf B}^{(-)}({\bf r}) = ( {\bf B}^{(+)}({\bf r})  )^{\da}$ and where the vector potential is
\bea \label{basic_field}    \nonumber
   {\bf A}^{(+)}({\bf r})  &=&     \mu_0 \, \int_0^{\infty} d \omega^{\, \prime} \int d^3 r' \, 
                                 \omega' \; \sqrt{ \frac{\hbar \epsilon_0}{\pi} \, 
                                 \epsilon_2({\bf r}',\omega') }
				 \\
				 &\times&
                    {\bf G}({\bf r},{\bf r}',\omega' ) \cdot  \hat{\bf f}({\bf r}',\omega')~ . 
\eea
   \noi
   Here the imaginary part of the complex permittivity is $\epsilon_2({\bf r},\omega)$.
   The dyadic Green's tensor ${\bf G}({\bf r},{\bf r}^{\, \prime},\omega)$ is the unique solution to the Helmholtz equation.
   Because the Helmholtz equation is linear, the associated Green's tensor can be
   written as a sum according to
\bea \label{G_tot}
  \bm{G}({\bf r},{\bf r}',\omega)=  \bm{G}^0({\bf r},{\bf r}',\omega)
+ \bm{G}^S({\bf r},{\bf r}',\omega) \, ,
\eea
    where $\bm{G}^0({\bf r},{\bf r}',\omega)$ represents the contribution of the direct waves from
    the radiation sources in an unbounded medium, which is vacuum in our case,
    and $\bm{G}^S({\bf r},{\bf r}',\omega)$ describes the scattering
    contribution of multiple  reflection waves from the body under consideration. The presence of the vacuum part
    $\bm{G}^0({\bf r},{\bf r}',\omega)$ in  Eq.(\ref{G_tot}) will in general give rise to divergences in the
    energy shifts to be calculated below. A renormalization prescription is therefore required.  We subtract the vacuum part
    $\bm{G}^0({\bf r},{\bf r}',\omega)$, i.e. we neglect possible finite corrections due to this vacuum
    subtraction like conventional vacuum-induced Lamb shifts. Since we, in the end, are going to consider
    CP forces all additive coordinate independent corrections to energy shifts will, anyway, not contribute.
    When we below refer to a renormalization prescription the procedure above is what we then have in mind.

   We now consider the Hamiltonian in \eq{H} and apply the well-known Weisskopf-Wigner theory 
   for the transitions $\alpha \rightarrow \beta$, where $E_{\alpha} > E_{\beta}$.
   The solution to the time-dependent Schr\"odinger equation in the rotating-wave approximation (RWA), i.e. applying
   $H' \approx H_{RWA}$, where
\bea
   H_{RWA} = - \sum_{\beta < \alpha}  |\alpha \rangle \langle \beta|  \, {\bm{\mu}}_{\alpha \beta}  \cdot  
                                      {\bf B}^{(+)}({\bf r}_A) + \textrm{h.c.} ~ ,
\eea
   \noi
   is then ($\omega_{\alpha} \equiv E_{\alpha}/\hbar$)
\bea  \nonumber
  | \psi(t) \rangle \, &=& \, c_{\alpha}(t) \, e^{- \, i \omega_{\alpha}  t} \, |\alpha \rangle \otimes |0 \rangle
                        \\
			&+&
                        \int d^3r \int_0^{\infty} d \omega ~  \sum_{m=1}^3 \sum_{\beta < \alpha}  \nonumber \\
			&\times&
			c_{m\beta}({\bf r}, \omega, t) \, e^{- \, i (\omega + \omega_{\beta}) t } 
		        | \beta \rangle \otimes | 1_m({\bf r}, \omega) \rangle ~ . ~~~
\label{psi_ansatz}
\eea
   \noi
   Here the initial state of the atom-field system is $|\alpha \rangle \otimes |0 \rangle$, where $|0 \rangle$ denotes the
   vacuum of the electromagnetic field and $| 1_m({\bf r}, \omega) \rangle = \hat{f}_m^{\da}({\bf r},\omega) \, |0 \rangle$
   is a one photon state. We also make use of the notation $\beta < \alpha$ in a $\beta$-sum to denote the condition $E_{\beta}<E_{\alpha}$.
   The coefficient $c_{\alpha}(t)$ is then determined by   
\bea   \label{cp_DL}
   \frac{d c_{\alpha}(t)}{dt}   =  \int_0^t  \sum_{\beta < \alpha } ~ dt^{\, \prime} ~ K^R_{\alpha \beta}(t-t^{\, \prime}) ~ c_{\alpha}(t^{\, \prime}) ~ , 
\eea
   \noi
   where $K^R_{\alpha \beta}(t)$ is the renormalized version of the kernel 
\bea  \label{K}
    K_{\alpha \beta}(t)  &=&    - \, \frac{1}{2 \, \pi} \, 
                      \int_0^{\infty} d\omega \, 
                      e^{- \, i  (\omega - \omega_{\alpha \beta})  t  } \, \Gamma_{\alpha \beta}({\bf r}_A, \omega) \,  . ~~ 
\eea
    \noi
    Here we have defined $\omega_{\alpha \beta} \equiv (E_{\alpha} - E_{\beta})/\hbar > 0$ as well as 
\bea   \nonumber
 && ~ \Gamma_{\alpha \beta}({\bf r}, \omega)  ~ =
                                             \\ \label{Gamma}
				             && 
 					     \frac{2 \, \mu_0}{\hbar} \, 
					     {\bm{\mu}}_{\alpha \beta} 
					     \cdot 
					     \mbox{Im}  [  \overrightarrow{\nabla} \times   
                                             {\bf G}({\bf r}, {\bf r}, \omega ) \times 
					     \overleftarrow{\nabla}  ] 
					     \cdot
					     {\bm{\mu}}_{\beta \alpha} \, , ~~~~
\eea
    \noi
    which is the spontaneous spin flip rate for the transition $\alpha \rightarrow \beta$ where $E_{\alpha} > E_{\beta}$.
    Assuming that the Markov approximation holds, i.e. memory effects can be discarded, we can make the substitution
    $c_{\alpha}(t')  \rightarrow c_{\alpha}(t)$ in \eq{cp_DL}. By extending the remaining time integral to infinite time
  and making use of the distributional identity
\bea
 \int_0^{\infty} dt \, e^{-i ( \omega - \omega_{\alpha \beta} ) t } =  
    \pi \, \delta(\omega - \omega_{\alpha \beta}) +  {\cal P} \frac{i}{ \omega_{\alpha \beta} - \omega } \, ,
\eea
   the solution to \eq{cp_DL} is then of the form 
 \bea    \nonumber
   c_{\alpha}(t)   &=&   c_{\alpha}(0) \, \exp \bigg \{  - \, 
                         [  ~ \frac{1}{2} \, \Gamma_{\alpha}({\bf r}_A, \omega_{\alpha})  
			 \\
			 &&\,  + ~ 
			 i  \, \delta \omega_{\alpha}({\bf r}_A, \omega_{\alpha})  ~ ] \, t  ~ \bigg \} \, . ~~
\label{eq:c_alpha} 
\eea
  \noi
  The spontaneous spin flip rate for an atom initially in the state $| \alpha \rangle$ is 
\bea
 \Gamma_{\alpha}({\bf r}, \omega_{\alpha}) = \sum_{\beta < \alpha } \, \Gamma^R_{\alpha \beta}({\bf r}, \omega_{\alpha \beta})~~,
\eea
 where $\Gamma^R_{\alpha \beta}({\bf r}, \omega_{\alpha \beta})$ is the renormalized version of the spontaneous spin flip rate in \eq{Gamma}. 
  In \eq{eq:c_alpha} we have also defined the frequency shift $\delta \omega_{\alpha}({\bf r}, \omega_{\alpha})$ by
\bea
 \delta \omega_{\alpha}({\bf r}, \omega_{\alpha}) = \sum_{\beta <\alpha}  \, \delta \omega^R_{\alpha \beta}({\bf r}, \omega_{\alpha \beta})~~,
\eea
where  again $ \delta \omega^R_{\alpha \beta}({\bf r}, \omega)$ is the renormalized version of
\bea   \label{delta_e}
 \delta \omega_{\alpha \beta}({\bf r}, \omega)   &=&    \frac{1}{2 \, \pi} ~   {\cal P}  \int_0^{\infty}
                                        \frac{ d \omega' }{\omega - \omega' } ~
                                        \Gamma_{\alpha \beta}({\bf r}, \omega') ~ ,
\eea
   \noi
   where $\Gamma_{\alpha \beta}({\bf r}, \omega)$ is given in \eq{Gamma}.
   Similar results to Eqs.(\ref{Gamma}) and (\ref{delta_e}) are also derived in Ref.\cite{fermani_06}.
   These latter results can, of course,  be obtained using conventional perturbation theory (see e.g. Ref.\cite{merzbacher_70}).

   In passing we observe  that \eq{Gamma}, i.e. the spontaneous spin flip rate, has been derived using the Heisenberg   	equations  of motion in Ref.\cite{rekdal_04}. Despite the fact that such an approach leads to a well-known mathematical
   inconsistencies due to operator ordering problems \cite{barnett_97},
   it gives, nevertheless, the same result.
   Similar results for the frequency shift for the more well-known case of an electric dipole can e.g. be found in
   Refs.\cite{scheel_99,vogel_06}.

  The principal integral in \eq{delta_e} can be further evaluated by means of well-known contour integral
  techniques due to the absence of poles in the upper complex frequency plane (see e.g. Refs.\cite{buhmann_04,barnett_97,vogel_06}). For the transitions $\alpha \rightarrow \beta$,
  where $E_{\alpha} > E_{\beta}$, we obtain  
\bea   
   \delta \omega_{\alpha}({\bf r}_A, \omega_{\alpha}) &=&   \delta \omega_{\alpha}^{(1)}({\bf r}_A, \omega_{\alpha})
                                                            +
                                                            \delta \omega_{\alpha}^{(2)}({\bf r}_A, \omega_{\alpha}) \; ,
\eea
   \noi
 where  $\delta \omega_{\alpha}^{(i)}({\bf r}, \omega_\alpha) = \sum_{\beta < \alpha}  \delta \omega_{\alpha \beta}^{(i)}({\bf r}, \omega_{\alpha \beta})$
   ($i=1,2$). Here we have defined the resonant contribution  
\bea   \nonumber
    &&~~~~~~~~~~~~~~~~~\delta \omega_{\alpha \beta}^{(1)}({\bf r}, \omega) ~ = 
    \\
    && 
    - \, \frac{\mu_0}{\hbar} \;
    {\bm{\mu}}_{\alpha \beta} \cdot \mbox{Re}  [  \overrightarrow{\nabla} \times   
      {\bf G}({\bf r}, {\bf r}, \omega )  
      \times \overleftarrow{\nabla}  ]  \cdot {\bm{\mu}}_{\beta \alpha} ~ , ~~~~~~~
\eea
    \noi
 and the off-resonant contribution 
\bea   \nonumber
     &&~~~~~~~~~~~~~~~~~   \delta \omega_{\alpha \beta}^{(2)}({\bf r}, \omega)  ~ =
     \\ \label{omega2} 
     && 
     \frac{\omega}{\pi}  \,  \frac{\mu_0}{\hbar} \,  \int_0^{\infty} d \omega' ~   
     \frac{ {\bm{\mu}}_{\alpha \beta} \cdot [  \overrightarrow{\nabla} \times   
            {\bf G}({\bf r}, {\bf r} , i \omega' )  
  	    \times \overleftarrow{\nabla}  ] \cdot {\bm{\mu}}_{\beta \alpha}
          }{ \omega^2 +  \omega'^2 } \, . ~~~~~~
\eea
     \noi

   In the RWA, the energy shift for the ground state in the Weisskopf-Wigner approach is, however, vanishing in the large time limit.
   On the other hand, applying the interaction $H' \approx H_{AWR}$,  where
\bea
    H_{AWR} = - \sum_{ \alpha > \beta}  |\alpha \rangle \langle \beta|  {\bm{\mu}}_{\alpha \beta}  \cdot  
                                        \hat{{\bf B}}^{(-)}({\bf r}_A) + \textrm{h.c.} ~ ,
\eea
   \noi
   rather than $H_{RWA}$ when considering the transitions $\beta \rightarrow \alpha$, 
   where $E_{\beta} < E_{\alpha}$, i.e. $\omega_{\beta\alpha} < 0$, 
   the decay rate vanishes but the energy shift is non-zero. It is given by
\bea
   \delta \omega_{\beta}({\bf r}_A, \omega_{\beta}) = \sum_{\alpha > \beta}  \, 
                                                      \delta \omega^R_{\beta \alpha}({\bf r}_A, \omega_{\beta \alpha}) \, , ~~~
\eea
  \noi
  where again $ \delta \omega^R_{\beta \alpha}({\bf r}, \omega)$ is the renormalized version Eq.(\ref{delta_e}).
     In the same fashion as above one may show that for the state $| \beta \rangle$, the frequency shift is
\bea   \label{delta_g_CIT}
    \delta \omega_{\beta}({\bf r}_A, \omega_{\beta}) =  \delta \omega_{\beta}^{(2)}({\bf r}_A, \omega_{\beta}) ~ ,
\eea
   where now
   $\delta \omega_{\beta}^{(2)}({\bf r}, \omega_{\beta}) = \sum_{\alpha > \beta}  \delta \omega_{\beta \alpha}^{(2)}({\bf r}, \omega_{\beta \alpha})$. For states such that $\omega_{\beta\alpha}<0$ there will be only off-resonant contributions when converting the $\omega$-integration in \eq{delta_e} to imaginary frequencies.
\begin{figure}[t]

\begin{picture}(0,0)(150,310)   

\includegraphics{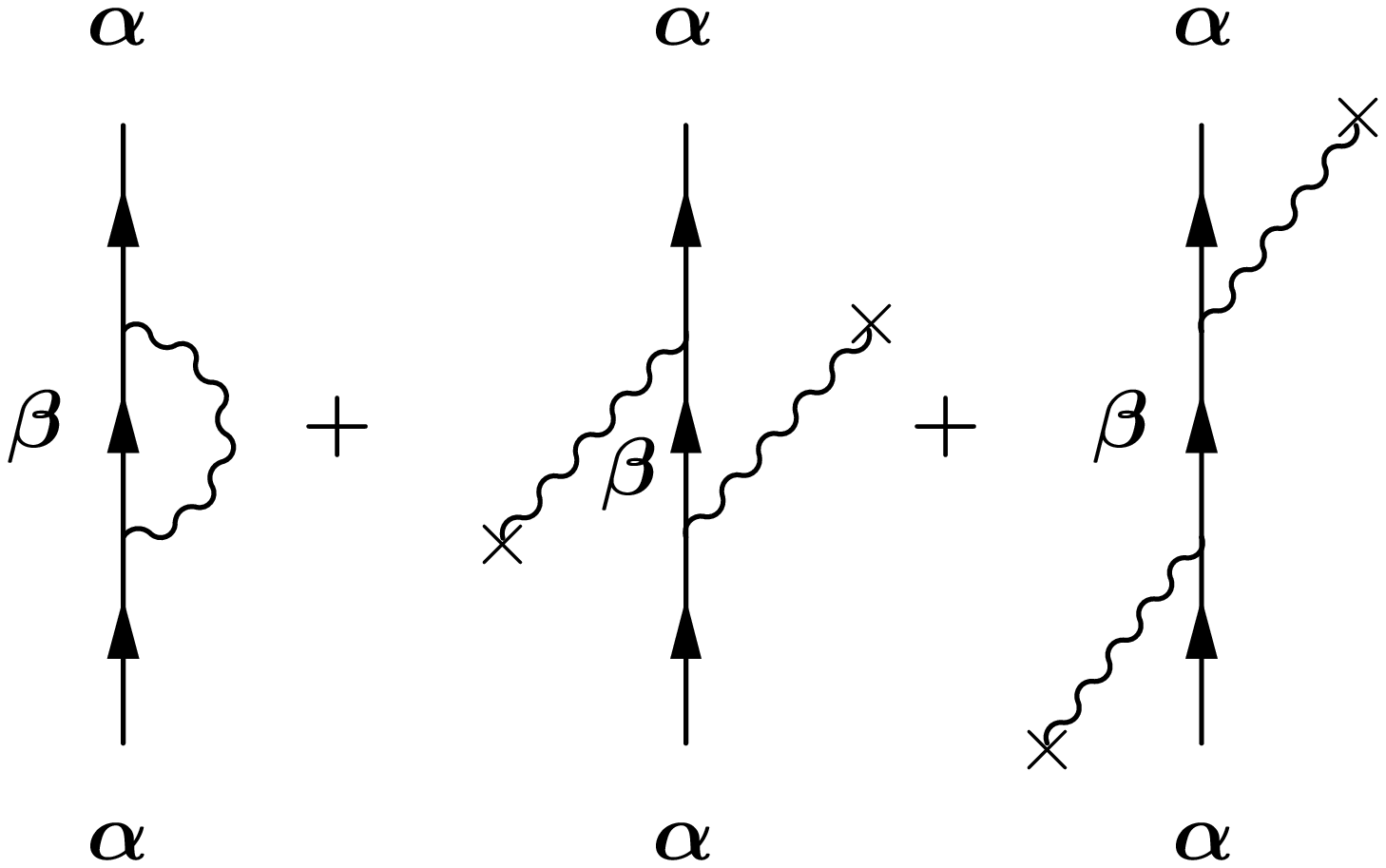}

\end{picture}

\vspace{5cm}

\caption{Feynman diagrams. From left to right:  the virtual process in vacuum, stimulated emission in the heath bath, and  the absorption-emission process in the heath bath, all leading to the same final state ${\bm{ \alpha}}$.}
\label{feynman_fig}

\end{figure}

From these discussions we now realize that, in general, it is not sufficient to make us of only  the effective interaction $H_{RWA}$ in the rotating-wave approximation. Instead we must use the complete interaction Hamiltonian $H'$ as given by Eq.(\ref{H_prime}).
This has the unfortunate  consequence that an exact analysis is not any more possible. The Weisskopf-Wigner theory can, however, easily be extended to a more general situation, at least when one limit oneself to self-energy corrections to at most second-order in perturbation theory (see e.g. Ref.\cite{merzbacher_70}). The energy shift for a given state of interest to us is then obtained by calculating the second-order energy shift using the complete interaction Hamiltonian Eq.(\ref{H_prime}). If we limit ourselves to perturbation theory an alternative procedure is to consider the coordinate-dependent energy shift and making use of linear response theory in a standard manner (see e.g. Ref.\cite{fetter_71}).  We are then interested in finding an expression for $ \langle \psi(t) |  H' | \psi(t) \rangle  =  \, _I \langle \psi(t) |H'_I(t) | \psi(t) \rangle_I$, where in the interaction picture and to lowest order in $H' $,

\bea   \label{int_pict}
    | \psi(t) \rangle_I = | \psi(t_0) \rangle +\frac{1}{i\hbar}\int_{t_0}^tdt'H'_I(t')| \psi(t_0) \rangle ~ .
\eea
Here we eventually consider the limit $t_0 \rightarrow -\infty$ for the initial time $t_0$. In terms of a retarded response function ${\cal D}_R(t,t')$ we can therefore write
\bea   \label{lin_respons}
 \langle \psi(t) |  H' | \psi(t) \rangle = \frac{1}{\hbar}\int_{-\infty}^{\infty}dt'{\cal D}_R(t,t') ~ ,
\eea
where 
\bea   \label{retarded}
i{\cal D}_R(t,t') = \Theta(t-t')\langle \psi(t_0) |  [H'_I(t) ,H'_I(t')]| \psi(t_0) \rangle ~,
\eea
and where we have assumed that $\langle \psi(t_0) |  H' | \psi(t_0) \rangle = 0$. The retarded Green's function Eq.(\ref{retarded}) is then directly related to the retarded Green's functions for our basic quantum fields ${\bf A}^{(+)}({\bf r})$, and its hermitian conjugated field, as given by Eq.(\ref{basic_field}), and can be calculated in a standard manner with a knowledge about the initial state $| \psi(t_0)\rangle$. The virtue of this formulation is that we now can allow for any quantum state of the radiation field, like a thermal background, of photons by making use of a  suitable random phases for a pure state of the radiation field.  One is then led to a consideration of so called real-time finite temperature propagators which has been considered elsewhere in great detail (see e.g. Refs.\cite{dolan_jackiw_74}). In the present paper we will, however, restrict us to an initial vacuum state of the radiation field and return to the case of a background of thermal photons elsewhere Ref.\cite{rek_ska_vas_09}. The energy shift then obtained will contain processes corresponding to photon absorption and emission, as well as stimulated emission, with a thermal background of photons. This physical picture is illustrated by means of the Feynman diagrams in Fig.\ref{feynman_fig}. The initial state of the atom is then assumed to be fixed which has been referred to as a constrained free energy of an atom immersed in heat bath Ref.\cite{barton_87}. 

For reasons of completeness we now write down the energy shift for a background of thermal photons, at a temperature $T$, including finite temperature spontaneous emission and absorption processes, valid to second-order in perturbation theory using the Feynman diagrams as indicated in Fig.\ref{feynman_fig}. We obtain $\delta \omega_{\alpha}({\bf r}, \omega_{\alpha},T) = \sum_{\beta }  \, \delta \omega^R_{\alpha \beta}({\bf r}, \omega_{\alpha \beta},T)$,
  where $ \delta \omega_{\alpha \beta}^R({\bf r}, \omega,T)$ is the renormalized version of
\bea   \label{delta_e_T}
 \delta \omega_{\alpha \beta}({\bf r}, \omega,T)   = ~~~~~~~~~~~~~~~~~~~~\nonumber \\    \frac{1}{2 \, \pi} ~   {\cal P}  \int_0^{\infty}
                                        \frac{ d \omega' }{\omega - \omega' } ~
 (1+ n(\omega'))\Gamma_{\alpha \beta}({\bf r}, \omega')  ~ +\nonumber \\
 \frac{1}{2 \, \pi} ~   {\cal P}  \int_0^{\infty}
                                        \frac{ d \omega' }{\omega + \omega' } ~
 n(\omega')\Gamma_{\alpha \beta}({\bf r}, \omega') ~\, , ~~~~~~~
\eea
   \noi
   where $\Gamma_{\alpha \beta}({\bf r}, \omega)$ again is as given in \eq{Gamma} and $n(\omega) = 1/(\exp(\hbar\omega/k_BT) -1)$ is the Planck black-body distribution. As the temperature $T\rightarrow  0$ in Eq.(\ref{delta_e_T}) we  obviously have a smooth limit (see in this context e.g. Ref. \cite{Lamoreaux_08}).
   Similar results to Eqs. (\ref{Gamma}) and (\ref{delta_e}) are also derived in Ref.\cite{henkel_99,gorza_06,fermani_06,buhmann_08} for electric dipole transitions.

\vspace{0.5cm}
\bc{
\section{Atom Near a Non-Magnetic Slab}
\label{sec_half}}
\ec

    Let us now consider the geometry as shown in \fig{geo_slab_fig}, i.e. a semi-infinite slab with finite thickness.
    The analytical continuation of the equal position Green's tensor \eq{G_slab} is then
\bea    \nonumber
 && \overrightarrow{\nabla} \times {\bf G}^S({\bf r},  {\bf r}, i \omega ) \times  \overleftarrow{\nabla}
                             =  
 \\  \label{dd_G_ac}
 && ~ \frac{1}{4 \pi} 
         \left  [ ~
                                                     \ba{rrr}
                                       I_{\|}(\omega)  ~ & 
                                       0  ~~~~ &  
                                       0  ~~~~
         \\                            0  ~~~~ & ~ 
                                       I_{\|}(\omega)  ~ &  
                                       0  ~~~~
         \\                            0  ~~~~ &     
                                       0  ~~~~ & ~  
                                       I_{\perp}(\omega)  ~
                                                       \ea
                                                \right ]  ~ , ~~~~
\eea
    where $\omega$ is real and
\bea   \nonumber
    I_{\|}(\omega)  &=&  \frac{1}{2} \,
             \int_{0}^{\infty}  d \lambda  \, \frac{\lambda}{\eta_0(\lambda, \omega)} \, e^{- \, 2 \, \eta_0(\lambda, \omega) \, z} \,
    \\ \label{I_11_og_22}
    &\times & \,    \left\{ \, - \, \frac{\omega^2}{c^2} \, {\cal C}_{N}(\lambda, \omega)  +  \eta_0^2(\lambda, \omega) \, {\cal C}_{M}(\lambda, \omega) \, \right\} \, ,
    \\ \nonumber 
    \\  \label{I_33}
    I_{\perp}(\omega)  &=& 
		 \int_{0}^{\infty}  d \lambda \, \frac{\lambda^3}{\eta_0(\lambda, \omega)} \,  
		 e^{- \, 2 \, \eta_0(\lambda, \omega) \, z}  \, {\cal C}_{M}(\lambda, \omega) \, . ~~~~~~~~~
\eea
   The scattering coefficients $C_{N}(\lambda,\omega)$ and $C_{M}(\lambda,\omega)$, 
   which are analytical continuations of Eqs. (\ref{C_N_33_A}) and (\ref{C_M_33_A}), are given by \cite{li_94} 
\bea  \label{C_N_33}
  {{\cal C}}_{N}(\lambda, \omega)  =     r_p(\lambda, \omega) ~ \frac{1 - e^{- \, 2 \, \eta(\lambda, \omega) \, h} }
                                                                     {1 - r_p^2(\lambda, \omega)  \; e^{- \, 2 \, \eta(\lambda, \omega) \, h}} \, ,
    \\ \label{C_M_33}
  {{\cal C}}_{M}(\lambda, \omega)  =     r_s(\lambda,\omega) ~ \frac{1 - e^{- \, 2 \, \eta(\lambda, \omega) \, h} }
                                                                    {1 - r_s^2(\lambda, \omega)  \; e^{- \, 2 \, \eta(\lambda, \omega) \, h}} \, ,
\eea
   \noi
   with the electromagnetic field polarization dependent Fresnel coefficients
\bea    \label{r_s}
  r_s(\lambda, \omega)  &=&  \frac{\eta_0(\lambda, \omega) - \eta(\lambda, \omega)}{\eta_0(\lambda, \omega) + \eta(\lambda, \omega)} \; , \;
  \\    \label{r_p}
  r_p(\lambda, \omega)  &=& 
                           \frac{\epsilon(i \omega) \, \eta_0(\lambda, \omega) - \eta(\lambda, \omega)}
                                {\epsilon(i \omega) \, \eta_0(\lambda, \omega) + \eta(\lambda, \omega)} \, , ~~ 
\eea
    \noi 
    which are analytical continuations of Eqs. (\ref{r_s_A}) and (\ref{r_p_A}).
    Here we have defined $\eta(\lambda, \omega)  =  \sqrt{k^2 \epsilon(i \omega) + \lambda^2}$, $\eta_0(\lambda, \omega) = \sqrt{k^2 + \lambda^2}$
    and $k=\omega/c$. Below we will mostly consider the large $h$ limit, i.e. $h$ is supposed to be large in comparison with any other length scale in the system.

    Due to causality, the complex dielectric function $\epsilon(\omega) = \epsilon_1(\omega) + i \epsilon_2(\omega)$
    will in general obey the Kramers-Kronig relations. Hence \cite{lifshitz_84}
\bea
  \epsilon( i \omega ) = 1 \, + \, \frac{2}{\pi} \, \int_0^{\infty} dx \, \frac{x \, \epsilon_2(x)}{x^2 + \omega^2} \, ,
\label{KK}
\eea
    where $\epsilon(i \omega)$ is a real function.
    Due to physically necessary conditions, $\epsilon_2(x) \geq 0$ for $x \geq 0$ (see e.g. Ref.\cite{lifshitz_84}).
    This guarantees, in general, that ${\bf G}^S({\bf r},  {\bf r}', i \omega )$ is real.

    The  Casimir-Polder force  obtained from any frequency shift in $\delta \omega_{\beta}({\bf r}, \omega_{\beta })$ is now given by
\bea
\label{eq:CP_Force}
    {\bf F}_{\beta}({\bf r},\omega_{\beta}) = -  { \bf \nabla} [ \, \hbar \, \delta \omega_{\beta}({\bf r}, \omega_{\beta })  \, ]~~,
\eea
which actually is independent of the additive and divergent contribution due to the free Green's function $\bm{G}^0({\bf r},{\bf r},\omega)$.

    Let us also limit our attention to ground state transitions $\beta \rightarrow \alpha$, where $E_{\beta} \leq E_{\alpha}$ or $\omega_{\beta\alpha} \leq 0$,
    in which case we may apply Eqs. (\ref{delta_g_CIT}) and (\ref{dd_G_ac}).
    Only the off-resonant term contribute in this case, i.e.
\bea  \nonumber
    \delta \omega_{\beta}(z, \omega_{\beta})  &=&  \frac{\mu_0}{4 \pi \hbar}\sum_{\alpha}
    \bigg \{  ( | \mu_x^{\beta \alpha} |^2  +  | \mu_y^{\beta \alpha} |^2   ) \, i_{\|}(z,\omega_{\beta \alpha})  
    \\
    &+& \; 
    | \mu_z^{\beta \alpha}  |^2  \, i_{\perp}(z,\omega_{\beta \alpha})  
    ~ \bigg \} \, , ~~
\label{Eg_exact}
\eea
    \noi
    where we have defined ( using $\rho=\|,\perp$ ) 
\bea  \label{i_gamma}
    i_{\rho}(z,\omega)  &=&   \frac{\omega}{\pi} \, \int_0^{\infty} 
                                      \frac{d\omega'}{\omega^2 + \omega'^2} \; I_{\rho}(\omega')  \, .
\eea

\vspace{0.5cm}

In the numerical analysis we have found it useful to make use of the following property of $i_{\rho}(z,\omega)$, i.e. we can write
\bea  \label{i_gamma_2}
    i_{\rho}(z,\omega)  &=&    \frac{1}{\pi}\, \left( \int_0^{\epsilon}
                                      \frac{dx}{1 + x^2} \; I_{\rho}(\omega x)  \right. \nonumber \\ &+& \left.\int_0^{1/\epsilon} 
                                      \frac{dx}{1 + x^2} \; I_{\rho}(\omega /x)\right)\, ,
\eea
with a natural choice $\epsilon = 1$. As a curiosity we remark that such a behavior of integrals under a modular transformation $x\rightarrow 1/x$  has been noticed before and found to be useful in e.g. quantum electrodynamics \cite{skagerstam95}.

\vspace{5mm}
\bc{
\subsection{Perfect conductor}
\label{sec:general}}
\ec

    The integrals \eq{i_gamma} can, in general, not be computed analytically. For a perfect conductor  ${{\cal C}}_{N}(\lambda)=1$ and $ {{\cal C}}_{M}(\lambda)=-1$
    for any $h$. The corresponding integrals can then be simplified. In this case, the integrals Eqs.(\ref{I_11_og_22}) and (\ref{I_33}) are related by
\bea
I_{\|}(\omega) =  \frac{1}{2}I_{\perp}(\omega) -  \frac{k^2}{2z} e^{- 2 kz }\, .
\eea

    For an atom in the state $| \beta \rangle$, where $E_{\beta} \leq E_{\alpha}$, it is straight forward to show that
    the frequency shift is ($k_{\beta \alpha} \equiv |\omega_{\beta \alpha}/c|= k_{ \alpha\beta}$):
\bea  \nonumber
  \delta \omega_{\beta}(z, \omega_{\beta}) &=&  \frac{\mu_0}{4 \pi  \hbar} \frac{1}{(2 z)^3}\sum_{\alpha } 
    \bigg \{  ( |\mu_x^{\beta \alpha}|^2  + |\mu_y^{\beta \alpha}|^2  )\tilde{i}_{\|}(k_{\alpha\beta } z)
    \\ 
   &+&  
    |\mu_z^{\beta \alpha}|^2   \tilde{i}_{\perp}(k_{ \alpha\beta} z) 
    ~ \bigg \} \,  ,
\label{Eg_PC}\eea
   \noi
   where the dimensionless integrals 
   are given by
\bea  \label{i_para}
   \tilde{i}_{\|}(x)  &=&  \frac{2x}{\pi} \,  \int_0^{\infty} \frac{d\xi}{(2x)^2 + \xi^2} \,  
                           e^{- \xi} \, \bigg ( \, \xi^2  +   \xi +   1 \, \bigg ) \, , ~~~~
   \\ \label{i_perp}
   \tilde{i}_{\perp}(x)  &=&  \frac{4x}{\pi} \,  \int_0^{\infty} \frac{d\xi}{(2x)^2 + \xi^2} \,  e^{- \xi} \, 
                              \bigg ( \, \xi  +   1 \, \bigg ) \, . ~~~~
\eea
    \noi
    This result is analogous to Eq.(10.87) in Ref.\cite{vogel_06}, which describes the attractive potential for an
    electric dipole outside a perfectly conducting plate. The magnetic frequency shift for the excited state is given in the Appendix
    \ref{A_app}, see \eq{Ee_exact}. We observe that the energy shift corresponding to \eq{Eg_PC} is independent of $\hbar$, 
    i.e. it may be derived purely in classical manner.

    For short distances (non-retarded limit), i.e. $k_{\alpha\beta } z \ll 1$, one may easily show that
    $\tilde{i}_{\perp}(k_{\alpha\beta } z) \approx 2 \, \tilde{i}_{\|}(k_{\alpha\beta } z) \approx 1$,
    in which case \eq{Eg_PC} is reduced to
\bea   \nonumber
 \delta \omega_{\beta}(z,\omega_{\beta})  \simeq 
                   \frac{\mu_0}{64 \pi \, \hbar} \, \frac{1}{z^3} ~~~~~~~~~~~~~~
		   \\  \label{omega_g_64}
		   \times 
		   \sum_{\alpha }
                   \bigg\{\,   | \mu_x^{\beta \alpha}  |^2   +  | \mu_y^{\beta \alpha} |^2     +   2 \, | \mu_z^{\beta \alpha} |^2 \, \bigg\} \, \nonumber \\ =
 \frac{\mu_0}{64 \pi \, \hbar z^3} \, \langle \beta| {\bm \mu}\cdot {\bm \mu} + {\mu}_z{\mu}_z |\beta \rangle \, , ~~~~~~
\eea
due to completeness of states.
    Numerical studies show that Eq.(\ref{omega_g_64}) actually a good approximation for $k_{\alpha\beta } z \lesssim 0.01$.  
    This result may also be obtained using a classical approach, e.g. the method of images for a magnetic moment (see e.g. Ref.\cite{jackson_75}) .
    The frequency shift in \eq{omega_g_64} corresponding to the repulsive Casimir-Polder force 
    ${\bf F}_{\beta}(z,\omega_{\beta})$ as given by Eq.(\ref{eq:CP_Force})  i.e. 
\bea   
   {\bf F}_{\beta}(z,\omega_{\beta}) &\simeq &   \frac{3\mu_0}{64 \pi z^4 } \,
 \langle \beta| {\bm \mu}\cdot {\bm \mu} + {\mu}_z {\mu}_z |\beta \rangle  \hat{\bf n}_z ~ . ~~~~~~
\eea 
    Here $\hat{\bf n}_z$ is a
    unit vector in the $z$-direction, i.e. normal direction to the plane of the conductor. With  $a=x,y,z$ and $\mu_a^{\alpha \beta} = g_S e\hbar  S_a^{\alpha \beta}/2m_e$, where $S_a^{\alpha \beta}=\langle \alpha| \hat{S}_a |\beta \rangle$ are dimensionless,  the frequency shift $\delta \omega_{\beta}(z,\omega_{\beta \alpha})$ as given by Eq.(\ref{omega_g_64}) can be re-written as 
\bea    
  \delta \omega_{\beta}(z,\omega_{\beta}) \simeq    \frac{g_S^2}{64 \pi \, \epsilon_0 \hbar z^3} \, 
                                           \, \big( \, \frac{e}{2} \, \lambda_e \, \big)^2 \, 
\langle \beta| {\bm S}\cdot {\bm S} + {S}_z{S}_z |\beta \rangle \, , ~~~~
\eea
    where $\lambda_e = \hbar/m_e c$ is the
    Compton wavelength.
    For electric dipole transitions, the corresponding frequency shift can be written in an analogous form where the
    magnetic dipole moment  ${\bm{\mu}}_{\alpha \beta}$ is replaced by the electric dipole moment
    ${\bf d}_{\alpha \beta} = e  \, {\bf r}_{\alpha \beta}$ (see e.g. Ref.\cite{vogel_06}), where
    ${\bf r}_{\alpha \beta}  = \langle \alpha | \hat{\bf r} | \beta \rangle$.
    Here ${\bf r }$ is the displacement vector, 
    whose order of magnitude is the Bohr radius $a_0$. As $( \lambda_e / a_0 )^2 \approx 10^{-6}$ we realize that
    magnetic dipole transition  CP forces are in general weak as compared to the corresponding electric dipole transitions CP forces.

   In the long distance (i.e. retarded) limit, corresponding to $k_{ \alpha\beta} z \gg 1$, we realize that 
   $\tilde{i}_{\perp}(k_{\alpha\beta} z) \simeq \tilde{i}_{\|}(k_{\alpha\beta} z) \simeq 2/(\pi \, k_{\alpha\beta} z)$.
   Numerical studies show that this is a good approximation for $k_{\alpha\beta} z \gtrsim 10$.  \eq{Eg_PC} is then reduced to 
\bea   \nonumber
 \delta \omega_{\beta}(z,\omega_{\beta})  &\simeq &
                \frac{\mu_0}{16 \pi^2 \, \hbar z^4} \,  \, 
		\\  \label{omega_g_16}
		\times
		\sum_{\alpha \neq \beta} \frac{1}{k_{\alpha\beta}} \, \bigg\{ \,   |\mu_x^{\beta \alpha} |^2  &+&  |\mu_y^{\beta \alpha}|^2 + |\mu_z^{\beta \alpha}|^2 \,  \bigg\} \, , ~~~~~
\eea
and the corresponding repulsive CP force will then have a $1/z^5$ dependence. Eq.(\ref{omega_g_16}) is an complete analogy with the famous Casimir-Polder energy shift for electric dipole transitions in the large distance limit \cite{casimir_polder_48}. The factor containing the sum over intermediate states in Eq.(\ref{omega_g_16}) is then related to the static polarizability of the atom considered. In our case the corresponding factor could similarly  be interpreted as a static magnetization of the atom.

\bc{
\subsection{Metal}
\label{sec:metal}}
\ec

    In vacuum, the Drude dispersion relation may be written (see e.g. Ref. \cite{brevik_05})
\bea
  \epsilon( i \omega ) = 1 + \frac{\omega_p^2}{\omega \, ( \omega + \nu \, )} ~~,
\label{eq:brevik}
\eea
   where for gold the plasma frequency is $\omega_p = 9.0$ eV and the relaxation frequency is $\nu = 35$ meV. 

Let us first consider the case $\nu = 0$ and hence $\epsilon( i \omega ) = 1 +\omega_p^2/\omega^2$.  In order to present analytical as well as numerical results  we now consider a two-level approximation ($\beta = g$ and $e=\alpha$) for an infinitely thick slab.  The magnitude of the ground state  CP force is
\bea
F_g(z,\omega_A) = - d\delta\omega_g(z,\omega_A)/dz\, ,
\label{eq:Fg}
\eea
 with $\omega_A \equiv \omega_e - \omega_g$. It is now convenient to explicitly perform the spatial $z-$derivative in equations Eq.(\ref{I_11_og_22}) and (\ref{I_33}). This enables us to define a  rescaled and dimensionless force quantity $F_M(z,\omega_A)\equiv F_M(k_Az)$ for magnetic moment transitions as
\bea
\label{eq:scaledCP}
F_M(k_Az) \equiv F_g(z,\omega_A)\frac{32\pi z^4}{\mu_0 (\mu_Bg_S)^2} ~~~~~~~~~~~~~ \nonumber \\
 =  (|S_x|^2 + |S_y|^2) {\bar i}_{\|}(k_Az) + |S_z|^2 {\bar i}_{\perp}(k_Az)~~,
\eea
where $S_a \equiv S^{ge}_a$, $a=x,y,x$, and $k_A = \omega /c$. Here we have defined the functions
\bea
\label{eq:ibar}
{\bar i}_{\rho}(k_Az) = \frac{k_Az}{\pi}\int_{0}^{\infty}\frac{d\xi}{(k_Az)^2+\xi^2}{\bar I}_{\rho}(\xi, \alpha k_Az)~~,
\eea
where $\rho =\|$ or $\perp$ and $\alpha \equiv \omega_p/\omega_A$, in terms of 
\bea
\label{eq:int_perp}
{\bar I}_{\perp}(\xi, \gamma) =  2^4\int_{0}^{\infty}dxx^3 e^{-2\sqrt{x^2+\xi^2}}~~~~~~  \nonumber \\
\times \frac{\sqrt{\gamma^2+\xi^2+x^2}-\sqrt{\xi^2+x^2}}{\sqrt{\gamma^2+\xi^2+x^2}+\sqrt{\xi^2+x^2}}\, ,~~~~~~
\eea
as well as
\bea
\label{eq:int_para}
{\bar I}_{\|}(\xi, \gamma) = 2^3\int_{0}^{\infty}dx x e^{-2\sqrt{x^2+\xi^2}} ~~~~~~~~~~\nonumber \\
\times \bigg\{ (\xi^2+x^2)\frac{\sqrt{\gamma^2 + \xi^2+x^2}-\sqrt{\xi^2+x^2}}{\sqrt{\gamma^2+\xi^2+x^2}+\sqrt{\xi^2+x^2}}~~~~~~\nonumber \\
+ \,\xi^2\frac{(\xi^2+\gamma^2)\sqrt{\xi^2+x^2}-\xi^2 \sqrt{\gamma^2+\xi^2+x^2}}{(\xi^2+\gamma^2)\sqrt{\xi^2+x^2}+\xi^2 \sqrt{\gamma^2+ \xi^2+x^2}}\bigg\}\,.
\eea

Even though the Eqs.(\ref{eq:int_perp}) and (\ref{eq:int_para}) appear complicated their asymptotic expansions can, however,  be obtained in a straightforward manner. If $k_Az \lesssim  1$ then, according to Eq.(\ref{eq:ibar}), only small $\xi$ contributes. If, in addition, $\alpha k_Az  \lesssim  1$ then
\bea
{\bar I}_{\perp}(\xi, \alpha k_Az ) \simeq (\alpha k_Az )^2\, , ~~~~~~~~~~~~ \nonumber \\ 
\, {\bar I}_{\|}(\xi, \alpha k_Az ) \simeq  (\alpha k_Az )^2 \bigg( \frac{1}{2} + \frac{2\xi^2}{(\alpha k_Az)^2+2\xi^2} \bigg) \, ,
\eea
where we refrain from writing down the higher order terms.

\begin{figure}[t]
\begin{picture}(0,0)(165,265)   

\includegraphics{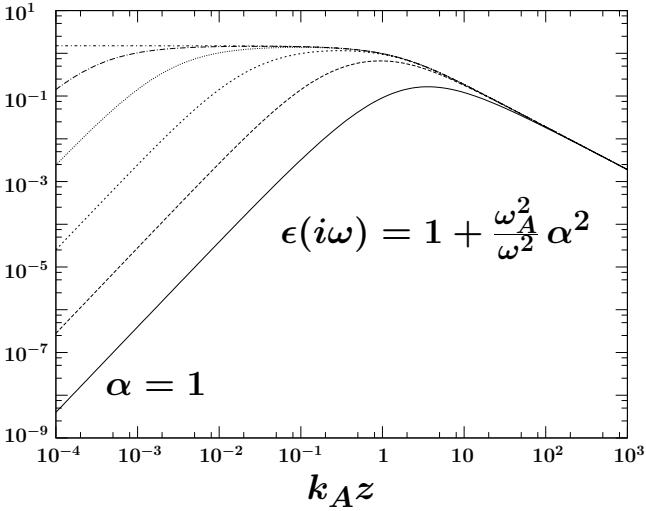}
\end{picture}
\vspace{7cm}
\caption{The  ground state dimensionless CP force $F_M(k_Az)$, as defined in the main text Eq.(\ref{eq:scaledCP}),  in the case of a dielectric constant $\epsilon (i\omega)= 1+\omega^2_{A}\alpha^2/\omega^2$ with $\alpha \equiv {\omega_p}/{\omega_A} $, as a function of $k_A z=\omega_Az/c$. For convenience we have chosen $|S_x|^2=|S_y|^2=|S_z|^2 = 1/4$. The curves correspond to $\alpha=1$ (lower solid curve), $\alpha=10,10^2,10^3,10^4$, and $\alpha =\infty$ (upper curve).}
\label{F_CP_1}
\end{figure}

If, on the other hand,  $\alpha k_Az \gtrsim 1$ we find that ${\bar I}_{\rho}(\xi,\alpha k_Az)$, for $a=\|, \perp$, does not depend on $\alpha$, and
\bea
{\bar I}_{\perp}(\xi, \alpha k_Az ) \simeq 2(3+6\xi+4\xi^2)e^{-2\xi}\, ,
\eea
as well as 
\bea
{\bar I}_{\|}(\xi, \alpha k_Az ) \simeq (3+6\xi+8\xi^2+8\xi^3)e^{-2\xi}\, .
\eea
Eqs.(\ref{eq:scaledCP}) and (\ref{eq:ibar}) then imply that if $k_Az$ and  $\alpha k_Az \lesssim  1$ then 
\bea
F_M(k_Az) \simeq  \left( (|S_x|^2 + |S_y|^2) \bigg\{ \frac{1}{2}+\frac{2}{2+\sqrt{2}\alpha}\bigg\}
\right. \nonumber \\ + |S_z|^2 
 \left) \frac{(\alpha k_Az)^2}{2}\right.\, ,
\eea
i.e. 
$F_g(z,\omega_A) \propto \alpha ^2/(k_Az)^2$. If, on the other hand, $\alpha k_Az \gtrsim 1$ but $k_Az$   sufficiently small, then  
\bea
F_M(k_Az) \simeq  \frac{3}{2}\left( |S_x|^2+ |S_y|^2+2|S_z|^2\right)\, ,
\eea
i.e. $F_g(z,\omega_A) \propto 1/(k_Az)^4$. With $\alpha k_Az \gg 1$ and $k_Az \gtrsim 10$,   we obtain
\bea
\label{eq:asym_fm}
F_M(k_Az) \simeq  \left( |S_x|^2+ |S_y|^2+|S_z|^2 \right)\frac{8}{\pi k_Az}\, ,
\eea
i.e. $F_g(z,\omega_A) \propto 1/(k_Az)^5$. In Fig.\ref{F_CP_1} the scaled and dimensionless CP force $F_M(k_Az)$ is presented  and the numerical calculations are in excellent agreement with the asymptotic expansions as given above. The choice for the matrix elements 
$|S_x|^2=|S_y|^2=|S_z|^2 = 1/4$ is not important and we do not in general find any essential difference with regards to $r$- and $p$-polarizations. We will therefore, for convenience, make use of the  same set of matrix elements in the numerical calculations throughout  the present paper. 
\begin{figure}[htb]
\begin{picture}(0,0)(150,253)   
\includegraphics{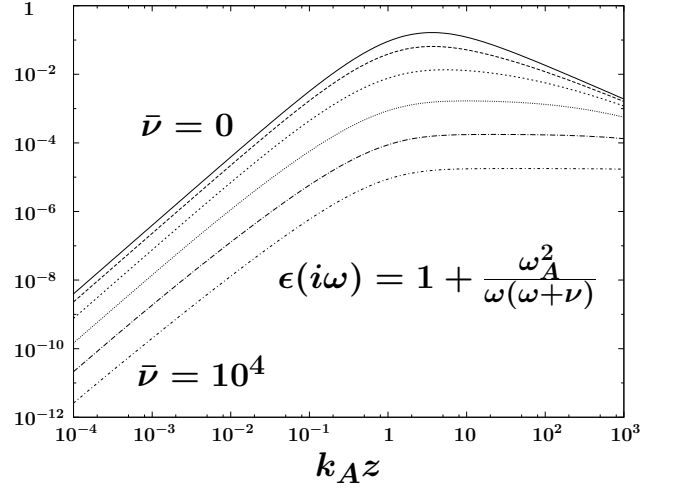}
\end{picture}
\vspace{6.5cm}
\caption{The  ground state dimensionless CP force $F_M(k_Az)$, as defined in the main text Eq.(\ref{eq:scaledCP}),  in the case of a dielectric constant $\epsilon (i\omega)= 1+\omega^2_{p}/\omega(\omega +\nu)$ with $\omega_{A}=\omega_{p}$,  as a function of $k_A z= \omega_Az/c$. For convenience we have chosen $|S_x|^2=|S_y|^2=|S_z|^2 = 1/4$.  With ${\bar \nu} \equiv {\nu}/{\omega_A} $ the curves correspond to ${\bar \nu=0}$ (upper solid curve),  ${\bar \nu} =1,10,10^2,10^3$, and ${\bar \nu}=10^4$ (lower curve).}
\label{gold_metal}
\end{figure}

Now returning to the Drude formula Eq.(\ref{eq:brevik}) with a finite $\nu$, we observe that the $\alpha \equiv \omega_p/\omega_A$-parameter in the Eqs.(\ref{eq:int_perp}) and (\ref{eq:int_para}) will be rescaled, i.e. $\alpha^2 \rightarrow  \alpha^2 \xi/(\xi+{\bar \nu})$ with ${\bar \nu} \equiv \nu/\omega_A$. For the parameters for gold mentioned above and for rubidium atoms with $\omega_A =2\pi\times 560$~kHz we then find that $\alpha \approx3.9 \cdot 10^9$ and ${\bar \nu} \approx 1.5 \cdot 10^7$ and one therefore expect that  we are close the perfect conductor limit. A numerical investigation verifies indeed that this is the case and $F_M(k_Az)$ is practically indistinguishable from the perfect conductor limit. For reasons of completeness we  show $F_M(k_Az)$ in Fig.\ref{gold_metal} for $\alpha =1$ for various values of ${\bar \nu}$. It is then clear that it is only for  $\alpha $ not to large that we are able to see the effect of dissipation due to  the presence of a finite $\nu$.

\begin{figure}[b]

\begin{picture}(0,0)(150,253)   

\includegraphics{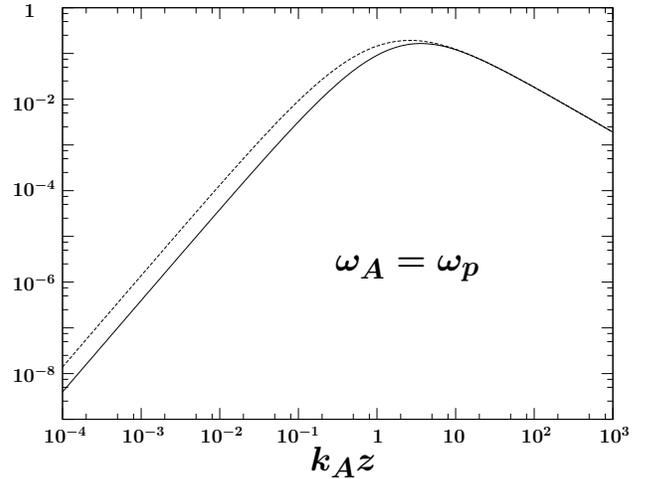}
\end{picture}
\vspace{6.5cm}
\caption{The  ground state dimensionless CP force $F_M(k_Az)$, as defined in the main text,  in the case of a dielectric constant $\epsilon (i\omega)= 1+\omega^2_{p}/\omega ^2$  as a function of 
$k_A z=\omega_Az/c$ (lower curve).  For convenience we have chosen $|S_x|^2=|S_y|^2=|S_z|^2 = 1/4$. The upper curve shows $F_M(k_Az)$ when including interband core electron transitions according to Ref.\cite{decca_et_al_2007} in terms of Eq.(\ref{eq:decca_2007}) with parameters as quoted in the main text.}
\label{gold_decca_et_al}

\end{figure}

More elaborate experimental investigations of the optical properties of gold  have been carried out by Decca et al. \cite{decca_et_al_2007} and by Svetovoy et al. \cite{svetovoy_2008}. In Ref.\cite{decca_et_al_2007} it was, in particular, argued that measured Casimir pressures could be used to find constraints on possible Yukawa couplings from non-standard particle physics. 
In this work a generalized 6-oscillator form of the dielectric, constant taking into account interband transitions of core electrons, was given in the form
\bea
\epsilon(i\omega) = 1+\frac{\omega_{p}^2}{\omega^2} + \sum_{j=1}^6 \frac{f_j}{\omega^2 +\omega_j^2+g_j\omega}\, ,
\label{eq:decca_2007}
\eea
with $\omega_p = 8.9$ eV and where the rest of the parameters $f_j$ and $g_j$ were fitted to known tabulated optical data. We have used the corresponding set of parameters in a calculation of rescaled and dimensionless force quantity $F_M(k_Az)$ in Eq.(\ref{eq:scaledCP}) in order to reveal the effect of a more precise description of the dielectric properties of gold. The results  are presented in Fig.\ref{gold_decca_et_al} and we conclude again that in general it will only be for $\omega _A  \simeq \omega_p$,  that details in the electromagnetic response properties of gold will be revealed in $F_M(k_Az)$ as defined by Eq.(\ref{eq:scaledCP}).

\vspace{0.5cm}

\begin{figure}[t]
\begin{picture}(0,0)(165,265)   

\includegraphics{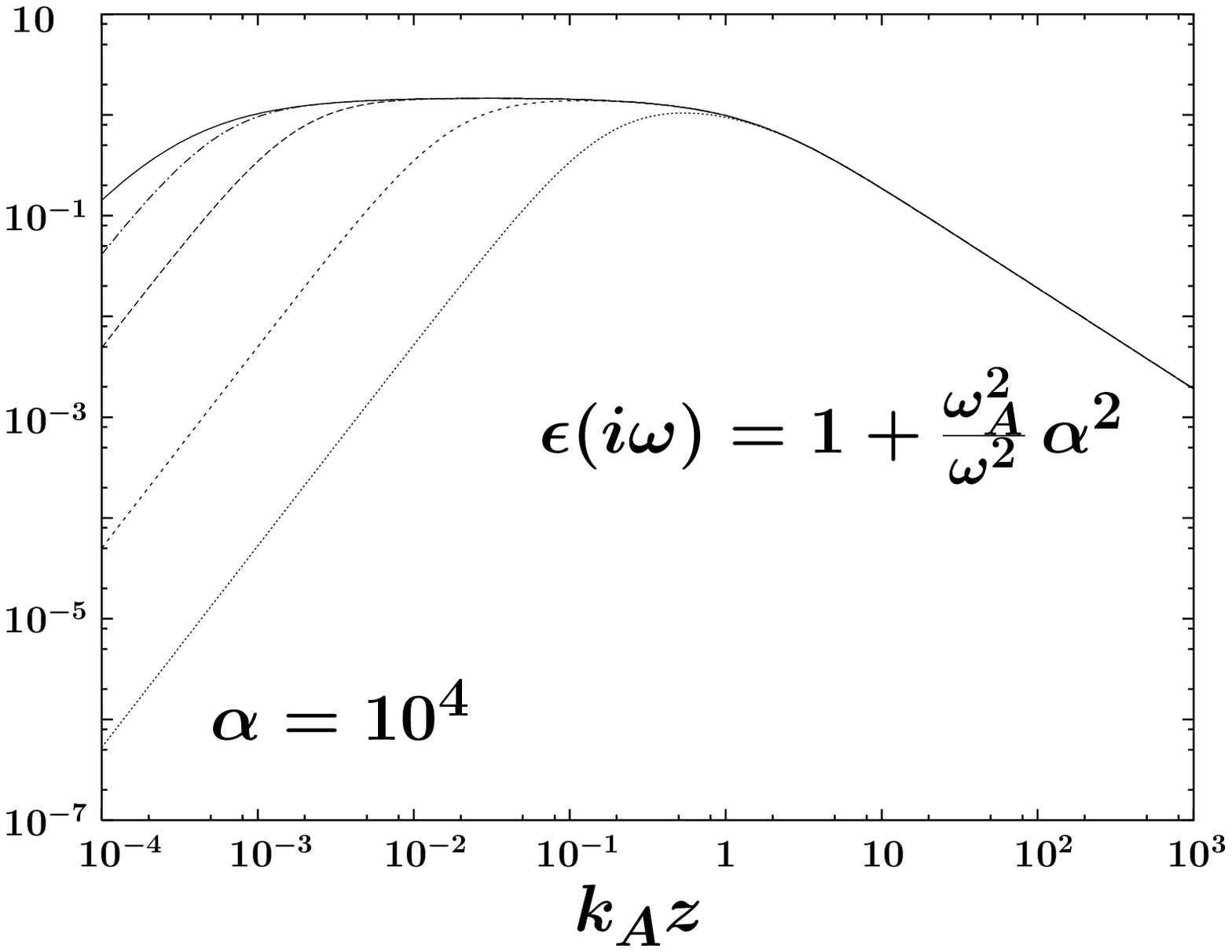}
\end{picture}
\vspace{7cm}
\caption{The  ground state dimensionless CP force $F_M(k_Az)$, corresponding to Eq.(\ref{Eg_exact}) in the main text, using a two-level approximation.  We make use of a dielectric constant $\epsilon (i\omega)= 1+\omega^2_{A}\alpha^2/\omega^2$, with $\alpha \equiv {\omega_p}/{\omega_A} $, as a function of $k_A z=\omega_Az/c$ for a fixed value $\alpha=10^4$ . For convenience we have chosen $|S_x|^2=|S_y|^2=|S_z|^2 = 1/4$. The curves correspond to $h/z= \infty$ (upper solid curve), $h/z=10^{-1},10^{-2},10^{-4}$, and $h/z=10^{-6}$ (lower curve). For $h/z= {\cal O}(1)$ we are already close to the infinite slab limit.}
\label{F_CP_h}
\end{figure}

In Fig.\ref{F_CP_h} we illustrate the dependence of the thickness $h$ of the slab on ground state dimensionless CP force $F_M(k_Az)$ in the two-level approximation, corresponding to Eq.(\ref{Eg_exact}), for  an almost perfect conductor ($\omega_p /\omega_A \gg 1$). We observe that for $h/z = {\cal O}(1)$ we are already close to the situation of a infinitely thick slab. We also notice that the dependence on the {\it geometrical} ratio $h/z$ is very similar to the dependence on the {\it material} property ratio $\omega_p /\omega_A$ for an infinitely thick slab as in Fig.\ref{F_CP_1}.

\begin{figure}[htb]

\begin{picture}(0,0)(150,253)   

\includegraphics{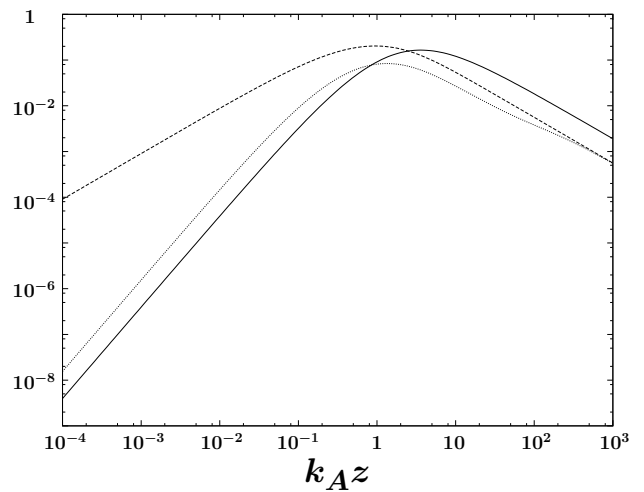}
\end{picture}
\vspace{6.5cm}
\caption{The  ground state dimensionless CP force $F_M(k_Az)$, as defined in the main text,  in the case of sapphire with a two-plateau dielectric function as given in the main text Eq.(\ref{eq:sapphire}).  For convenience we have chosen $|S_x|^2=|S_y|^2=|S_z|^2 = 1/4$. The upper dashed curve shows $F_M(k_Az)$ when $\omega_A = ck_A = 2\pi\times 560$ kHz. The middle dotted curve corresponds to $\omega_A = \omega_p$ for sapphire. For comparison we also give $F_M(k_Az)$ with a dielectric constant $\epsilon (i\omega)= 1+\omega^2_{p}/\omega ^2$ with $\omega_A=\omega_p$ (lower solid curve).}
\label{fig_sapphire}

\end{figure}

\vspace{0.5cm}
\bc{
\subsection{Dielectric medium}
\label{sec:diel}}
\ec
%
%
%
    As an example of a dielectric media we consider sapphire for which the electric dipole induced CP forces has been considered in Ref.\cite{antezza_2004}.
We find that the typical  "two-plateau" behavior of $\epsilon(i\omega)$ of sapphire as obtained numerically in Ref.\cite{antezza_2004} can be parameterized in the following form
\bea
\label{eq:sapphire}
\epsilon(i\omega) = 1+\frac{\omega_{p1}^2}{\omega^2 +\omega_1^2}+ \frac{\omega_{p2}^2}{\omega^2 +\omega_2^2}\, ,
\eea
with the parameters $\omega_{p1}=0.16$ eV, $\omega_{p2}=30.8$~eV, $\omega_{1}=0.07$ eV, $\omega_{2}=20.8$ eV,  where $\omega$ is given in units of eV. In Fig.\ref{fig_sapphire}
we illustrate the behavior  of the ground state two-level rescaled force $F_M(k_Az)$ where we now observe that $F_g(z,\omega_A) \propto 1/(k_Az)^3$ even in the case rubidium
atoms with a very small $\omega_A =2\pi\times 560$ kHz as compared to other parameters in Eq.(\ref{eq:sapphire}). This change of slope in $F_M(k_Az)$ as a function of $k_Az$
as compared to a metal is due to the fact that {\it all} the parameters in Eq.(\ref{eq:sapphire}) are large as compared to $\omega_A$. Indeed, using the dielectric constant Eq.(\ref{eq:sapphire})
and Eq.(\ref{eq:scaledCP}) as well as Eq.(\ref{eq:ibar}), one finds that if $k_Az \ll  1$ then ${\bar i}_{\rho}(k_Az) \simeq k_Az$ for $\rho =\|$ or $\perp$.

\vspace{5mm}

\bc{
\subsection{Superconductor}
\label{sec:super}}
\ec

    As the total current density is assumed to respond linearly and locally to the electric field,
    the dielectric function $\epsilon(\omega)$ at zero temperature can be written in the form
\bea \label{eps_j_2}
  \epsilon(\omega) = 1 - \frac{\sigma_2(\omega)}{\epsilon_0 \, \omega} + i \, \frac{\sigma_1(\omega)}{\epsilon_0 \, \omega} \, .
\eea
     Here $\sigma(\omega) \equiv \sigma_1(\omega) + i \sigma_2(\omega)$ is the, in general frequency dependent, complex optical conductivity.
     A detailed and often used description of the electrodynamic properties of superconductors  was developed by Mattis-Bardeen \cite{mattis_58}, and independently by
    Abrikosov-Gor'kov-Khalatnikov \cite{gorkov58},  based on the weak-coupling BCS theory of superconductors. 
    In the clean limit, i.e. $l \gg \xi_0$, where $l$ is the electron mean free path and $\xi_0$ is the coherence
    length of a pure material, the complex conductivity, normalized to $\sigma_n \equiv \sigma_1(T=T_c)$,
    can be expressed in the convenient form \cite{klein_94}
\bea  
  \frac{\sigma(\omega)}{\sigma_n} &=& \int_{\Delta - \hbar \omega}^{\Delta} \frac{dx}{\hbar \omega} \, 
        \, g(x)
   \label{sigma_klein}
 \, ,  
\eea
   \noi
   where
\bea
 g(x) = \frac{x^2 + \Delta^2 + \hbar \omega \, x }{u_1(x) u_2(x)} \, .
\eea
\begin{figure}[h]

\begin{picture}(0,0)(150,253)   

\includegraphics{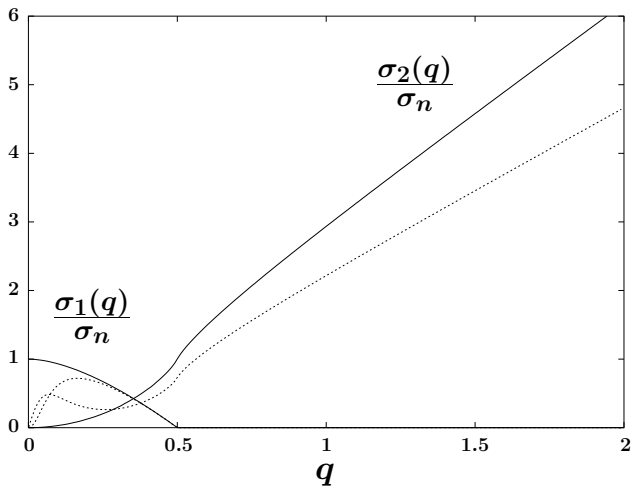}
\end{picture}
\vspace{6.5cm}
\caption{The complex conductivity according to the Mattis-Bardeen theory. The variable $q$ is given by $q\equiv \Delta/\hbar\omega$. 
We also observe that $\sigma_1(\omega) = 0 $ if $q \ge 1/2$. The solid curves correspond to an infinite relaxation time $\tau$. The dotted curves correspond to niobium  for which we choose $13.61=\hbar/\tau\Delta$, which corresponds to a mean free path of $9$ nm.}
\label{klein_chang_fig}

\end{figure}
 Here we have defined the functions
\bea
u_1(x) = \left\{ \begin{array}{ll }\mbox{sgn} (x)\sqrt{x^2 - \Delta^2} & \mbox{~~if  $|\Delta|$} < x ~, \\
 \\
-i\sqrt{\Delta^2 - x^2 } & \mbox{~~if $|\Delta|$} > x ~, \end{array} \right. \, 
\eea
as well as
\bea
u_2 (x) = \sqrt{(x + \hbar \omega)^2 - \Delta^2}\, . 
\eea
   Here, the well-known BCS  superconducting energy gap $\Delta$ is given by 
$\Delta = 3.53 \, k_B T_c/2$ in terms of a critical temperature $T_c$ \cite{fetter_71}.
 In Fig.\ref{klein_chang_fig} we plot $\sigma_1(\omega)$ and $\sigma_2(\omega)$ as functions of the dimensionless variable $q\equiv \Delta/\hbar\omega$. Since the low-frequency properties of $\epsilon(i\omega)$ will dominate the contribution to CP forces, it is important to notice that Eq.(\ref{sigma_klein}) leads to the asymptotic behavior $\sigma_2(\omega)/\sigma_n \rightarrow \pi\Delta/\hbar\omega$ as $\hbar\omega/\Delta \rightarrow 0$. The dielectric function $\epsilon(\omega)$ will therefore have a pole of second-order at $\omega =0$. 

The Kramers-Kronig relation Eq.(\ref{KK}) has then to be modified according to a standard procedure \cite{lifshitz_84} with the result
\bea
  \epsilon( i \omega ) = 1 \, + \frac{\omega_{sp}^2}{\omega^2} \, + \, \frac{1}{\pi} \, P\int_{-\infty}^{\infty} dx \, \frac{x \, \epsilon_2(x)}{x^2 + \omega^2} \, ,
\label{KK2}
\eea
where we have defined $\omega_{sp}^2 \equiv \pi\sigma_n\Delta/\epsilon_0\hbar$. For niobium $\hbar\omega_{sp} = 2.4$ eV which can be compared to $\hbar\omega_{p} = 9.0$ eV for gold.

As shown by Anderson and others \cite{anderson_59,abrikosov_58}, the presence of non-magnetic impurities,
    which we only consider in the present paper, will not modify the superconducting energy gap as given by $\Delta$.
    The complex conductivity will, however, in general be modified due to the presence of such impurities. In terms of the normal conducting state  relaxation time $\tau$,  due to the presence of non-magnetic impurities, one can write \cite{chang_89}
\bea  \nonumber
  \frac{\sigma_2(\omega)+i\sigma_1(\omega)}{\sigma_n}  = ~~~~~~~~~~~~~~~~~~~~~~~~~~~~
\\ \nonumber
  \frac{1}{2\omega\tau} \int_{\Delta - \hbar \omega}^{\Delta} dx  ~ 
      \bigg (  \frac{g(x) + 1}{u_2 - u_1 + i \hbar/\tau}  - 
                        \frac{g(x) - 1}{u_2 + u_1 - i \hbar/\tau}   \bigg )~~ 
  \\   
  -   \frac{1}{2\omega\tau}\int_{\Delta}^{\infty} dx ~   
   \bigg (  \frac{g(x) - 1}{u_2 + u_1 - i \hbar/\tau}   + 
                      \frac{g(x) - 1}{u_2 + u_1 + i \hbar/\tau}   \bigg ) \, . ~~~
\label{sigma_chang} 
\eea
For niobium we choose $\tau$ such that $\hbar/ \tau \Delta=\pi\xi_0 /l = 13.61$, corresponding to the
     experimental coherence length $\xi_0 = 39 $ nm and the mean free path $l(T \approx 9\, \mbox{K})=v_F\tau = 9$ nm \cite{pronin_98}. The low-frequency limit of Eq.(\ref{sigma_chang}) can be calculated in a straightforward manner and we find that
\bea
  \frac{\sigma_2(\omega)}{\sigma_n} \rightarrow \pi \frac{\Delta}{\hbar \omega} \left( 1 -f(\frac{\hbar}{\Delta\tau})\right)\, ,
\label{changasymp}
\eea
as $\hbar\omega/\Delta \rightarrow 0$, where we have defined the real-valued function 
\bea
  f(x)= 2\frac{\ln \left( x/2+\sqrt{(x/2)^2-1}~ \right)}{\pi\sqrt{(x/2)^2-1}} \, .
\label{ffunction}
\eea
We notice that in the presence of non-magnetic impurities Eq.(\ref{changasymp}) we still obtain a $1/\omega$ singularity in $\sigma(\omega)$ for small $\omega$.
In Fig.\ref{klein_chang_fig} we illustrate the $\omega$-dependence of $\sigma(\omega)$ in the clean limit ($\tau \rightarrow \infty$) as well as for niobium. Numerical investigations show that  the asymptotic formula Eq.(\ref{changasymp}) actually describes $\sigma_2(\omega)$ for a wide range of frequencies. We also notice that for any finite $\tau/\hbar\Delta \neq 0$  we have the asymptotic behavior $\sigma(\omega) \rightarrow 0$ as $\hbar\omega/\Delta \rightarrow \infty$. The discussion in this Section reveals that for superconductors, we are led back to the considerations for a metal as discussed in Section \ref{sec:metal},  apart from the continuum contribution in Eq.(\ref{KK2}). A closer of investigation shows that its only for frequencies $\omega_A  \gtrsim \Delta/\hbar$ that the CP force will be sensitive to this continuum contribution. For such frequencies, however, the Cooper pairs of the superconductor start to break up and we are again led back to a metal.
\vspace{0.5cm}

\begin{figure}[htb]

\begin{picture}(0,0)(150,253)   

\includegraphics{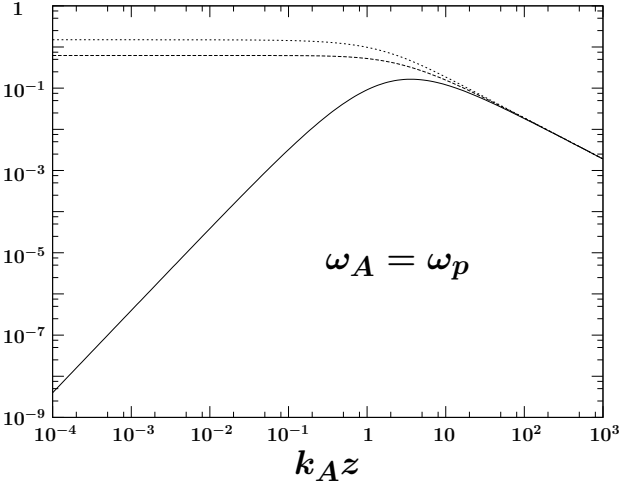}
\end{picture}
\vspace{6.5cm}
\caption{The attractive ground state dimensionless CP force $F_E(k_Az)$, as defined in the main text Eq.(\ref{eq:scaled_E_CP}) for electric dipole transitions. For convenience we have chosen $ |{\hat d}_x|^2=|{\hat d}_y|^2=|{\hat d}_z|^2=1/4$.   With a dielectric constant $\epsilon (i\omega)= 1+\omega^2_{p}/\omega ^2$,  the middle curve shows $|F_E(k_Az)|$ as a function of 
$k_A z= \omega_Az/c $ with $\omega_A= \omega_p$ and, for comparison, the lower curve corresponds to $F_M(k_Az)$ (repulsive) for $\omega_A= \omega_p$ and $|S_x|^2=|S_y|^2=|S_z|^2 = 1/4$.   The upper curve shows the same $F_M(k_Az)$ as well $|F_E(k_Az)|$ in the perfect conductor limit $\omega_p \rightarrow \infty$. }
\label{dual_case}

\end{figure}

\begin{figure}[htb]

\begin{picture}(0,0)(150,253)   

\includegraphics{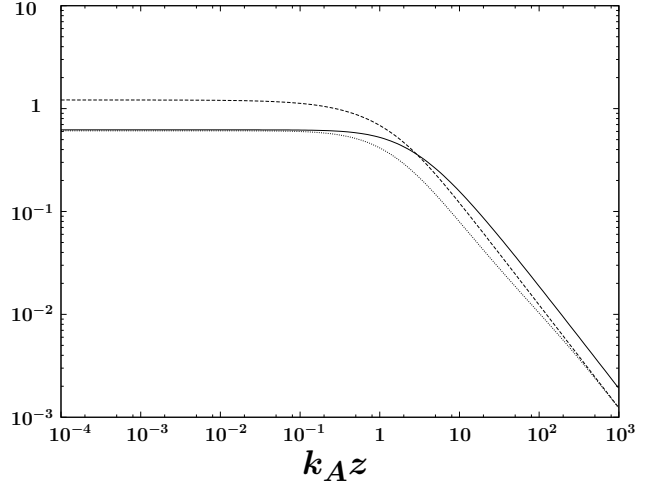}
\end{picture}
\vspace{6.5cm}
\caption{The absolute value $|F_E(k_Az)|$ of the  attractive ground state dimensionless CP force $F_E(k_Az)$, as defined in the main text Eq.(\ref{eq:scaled_E_CP}) for electric dipole transitions with $|{\hat d}_x|^2=|{\hat d}_y|^2=|{\hat d}_z|^2=1/4$.
The upper dashed curve gives $|F_E(k_Az)|$ for the rubidium transition  $\omega_A = ck_A=2\pi\times 560$ kHz 
and in the case of  sapphire with the two-plateau dielectric constant as given in the main text Eq.(\ref{eq:sapphire}).
For comparison the middle solid curve shows $|F_E(k_Az)|$ for a dielectric constant $\epsilon (i\omega)= 1+\omega^2_{p}/\omega ^2$ with  $\omega_A = \omega_p $.  The lower dotted curve shows $|F_E(k_Az)|$ in the case of sapphire but now with $\omega_A = 9.0$ eV. }
\label{fig_sapphire_dual_case}

\end{figure}
\bc{
\section{FINAL REMARKS}
\label{sec:summary}}
\ec
%
%

    In addition to the interaction Hamiltonian $H' \equiv -{\bm{\mu} } \cdot {\bf B}$ as given Eq.(\ref{H_prime}) we may also a consider  the Power-Zienau-Wooley \cite{power_59,woolley_71} Hamiltonian $H' \equiv -{\bf d } \cdot {\bf E}$ in order to describe electric dipole transitions.  For the semi-infinite half space geometry as considered in Section \ref{sec_half}  one can now easily transfer the results for magnetic transitions to electric dipole transitions according to the observations in Appendix \ref{A_app}. This has already been alluded to in Section \ref{sec:general}. There is, however, a subtle difference due to the role of the scattering coefficients $C_{N}(\lambda,\omega)$ and $C_{M}(\lambda,\omega)$ in Eqs.(\ref{I_11_og_22}) and (\ref{I_33}) which, for electric dipole interactions, has to be interchanged. We then define a dimensionless and rescaled attractive ground state CP force  $F_E(k_Az)$ in the two-level approximation  analogous to Eq.(\ref{eq:scaledCP}), i.e.
\bea
\label{eq:scaled_E_CP}
F_{E}(k_Az) \equiv F^{E}_g(z,\omega_A)\frac{32\pi z^4}{\mu_0 |{\bf d}|^2} ~~~~~~~~~~~~~~~~~~ \nonumber \\
=  (|{\hat d}_x|^2 + |{\hat d}_y|^2) {\bar i}^E_{\|}(k_Az) + |{\hat d}_z|^2 {\bar i}^E_{\perp}(k_Az) ~,\nonumber \\
\eea
where $F^{E}_g(z,\omega_A)$ is the ground state CP force induced by  electric dipole transitions and $|{\hat d}_a|^2 \equiv |d_a|^2/|{\bf d}|^2$, $a=x,y,z$. The functions ${\bar i}^E_{\perp}$ and ${\bar i}^E_{\|}$ correspond to Eq.(\ref{eq:ibar}) with the only difference that the scattering coefficients $C_{N}(\lambda,\omega)$ and $C_{M}(\lambda,\omega)$ have been interchanged. Similar to the analysis in Section \ref{sec:metal}, i.e. for a metal, and for $k_Az$ and $\alpha k_Az \lesssim 1$, one may show that 
\bea
\label{eq:E_integrals}
{\bar i}^E_{\|}(k_Az)\simeq -\frac{3}{2}\frac{\alpha}{\sqrt{2}+\alpha }~ \, , ~{\bar i}^E_{\perp}(k_Az)\simeq -\frac{3\alpha}{\sqrt{2}+\alpha }~~.
\eea
If, on the other hand, $\alpha k_Az \gg 1$ and $k_Az \gtrsim 10$   we obtain, apart from a minus sign, the same asymptotic expansion as in
Eq.(\ref{eq:asym_fm}) by replacing $|S_a|^2$ with $|{\hat d}_a|^2$ for $a=x,y,z$. In Fig.\ref{dual_case} we illustrate these limits for $F_{E}(k_Az)$ where we observe that it is only for magnetic moment transitions we obtain a behavior $F_g(z,\omega_A) \propto \alpha ^2/(k_Az)^2$ for small sufficiently $k_Az$.

In Fig.\ref{fig_sapphire_dual_case} we show the attractive CP force $F_E(k_Az)$ for sapphire as well as for a metal and for different frequencies $\omega_A$. We find the same qualitative features in all cases presented. Indeed, the discussion  as in Section \ref{sec:diel}, but for sapphire, reveals that it is only for magnetic transitions we obtain a behavior $F_g(z,\omega_A) \propto 1/(k_Az)^2$ or $1/(k_Az)^3$  for sufficiently small $k_Az$ as presented in Fig.\ref{fig_sapphire}.

\vspace{0.5cm}

\begin{center}ACKNOWLEDGEMENTS
\end{center}

\vspace{0.5cm}

     This work has been supported in part by the Norwegian University of Science and Technology (NTNU)
     and the Norwegian Research Council (NFR). One of the authors (B.-S.S.) wishes to thank Professors I.H. Brevik, K. Fossheim and
     J.S. H\o ye for various and fruitful illuminating comments and
Professor Frederik G.  Scholtz for a generous hospitality during a joint NITheP and Stias (S.A) workshop in 2009 when the present work was finalized. Members of the Norwegian network Complex,  in particular T.H. Johansen, are also acknowledged for their generous inspiration during the course of the present work. The constructive remarks from  anonymous
referee are also gratefully acknowledged.

\vspace{0.5cm}

\appendix

\section{Green's tensor for a Slab}
\label{A_app}

   The equal position scattering Green's tensor for the geometry as shown in \fig{geo_slab_fig}, i.e. a semi-infinite half space,
   is ($z \geq 0$ and $\omega$ is real)
\bea
        && \overrightarrow{\nabla} \times {\bf G}^S({\bf r} , {\bf r}  ,  \omega ) \times  \overleftarrow{\nabla}
                             ~ =  \nonumber
        \\ \nonumber
        && \frac{i}{8 \pi} \, \frac{\omega^2}{c^2}  \,   
	\int_{0}^{\infty}d \lambda \frac{\lambda}{\eta_0(\omega)} ~ e^{ i  2  \eta_0(\omega) \, z} 
        \\ \nonumber 
	&& \times ~~ \bigg \{ \; {\cal C}_{N}(\lambda) 
         \left  [ 
                                                     \ba{rrr}
                                       1   & 
                                       0   &  
                                       0  
         \\                            0   &  
                                       1   &  
                                       0  
         \\                            0   &     
                                       0   &   
                                       1  
                                                       \ea
                                                \right ]  ~~~~~
        \\  \label{G_slab}
	&& +  ~    {\cal C}_{M}(\lambda) \, \frac{c^2}{\omega^2}  
         \left  [ 
                                                     \ba{rrr}
                                       - \, \eta_0^2(\omega)   & 
                                       0  ~~~~~  &  
                                       0  ~~
         \\                            0  ~~~~~  &  
                                       - \, \eta_0^2(\omega)   &  
                                       0  ~~
         \\                            0  ~~~~~  &     
                                       0  ~~~~~  &  
                                       2 \, \lambda^2  
                                                       \ea
                                                \,   \right ]  \, \bigg \} ~ . ~~~~~~~  
\eea
   \noi
   This equation is valid for a finite thickness $h$ of the slab, i.e. it is more general than the analogous 
   Eq. (G1) in Ref. \cite{buhmann_04}.  
   Note that the analytical continuation of \eq{G_slab} is \eq{dd_G_ac}.
   The scattering coefficients $C_{N}(\lambda)$ and $C_{M}(\lambda)$ are given by \cite{li_94} 
\bea  \label{C_N_33_A}
  {{\cal C}}_{N}(\lambda)  =     r_p(\lambda) ~ \frac{1 - e^{i  2 \, \eta(\lambda)  h} }
                                                                     {1 - r_p^2(\lambda)  e^{i  2  \eta(\lambda) \, h}} \, ,
    \\ \label{C_M_33_A}
  {{\cal C}}_{M}(\lambda)  =     r_s(\lambda) ~ \frac{1 - e^{i 2  \eta(\lambda) \, h} }
                                                                    {1 - r_s^2(\lambda)  e^{i  2  \eta(\lambda) \, h}} \, ,
\eea
   \noi
   with the electromagnetic field polarization dependent Fresnel coefficients
\bea  \label{r_s_A}
  r_s(\lambda)  &=&  \frac{\eta_0(\lambda) - \eta(\lambda)}{\eta_0(\lambda) + \eta(\lambda)} \; , \;
  \\  \label{r_p_A}
  r_p(\lambda)  &=& 
                           \frac{\epsilon(\omega) \, \eta_0(\lambda) - \eta(\lambda)}
                                {\epsilon(\omega) \, \eta_0(\lambda) + \eta(\lambda)} \, . ~~ 
\eea
    \noi 
    Here we have defined
\bea
 \eta(\lambda)  =  \sqrt{k^2 \epsilon(\omega) - \lambda^2}\, ,
\eea
and 
\bea
\eta_0(\lambda) = \sqrt{k^2 - \lambda^2}\, ,
\eea
as well as $k=\omega/c$.

    For the case of electric dipole transitions, the equal position scattering Green's tensor ${\bf G}^S({\bf r}  ,  {\bf r}  ,  \omega )$
    rather than $\overrightarrow{\nabla} \times {\bf G}^S({\bf r}  ,  {\bf r}  ,  \omega ) \times  \overleftarrow{\nabla}$ is
    needed. This Green's tensor can be obtained from \eq{G_slab} by the substitution ${{\cal C}}_{N}(\lambda) \leftrightarrow {{\cal C}}_{M}(\lambda)$
    and multiplying with $c^2/\omega^2$. In addition there will be a $\omega ^2$ factor in the calculation of a transition rate $\Gamma$ due to the presence of the electric field in the electric dipole Hamiltonian.

\vspace{0.1cm}

\section{Perfect Conducting Slab}
\label{B_app}

   For an excited atom above a perfect conducting slab as shown in \fig{geo_slab_fig}, the frequency shift is 
\bea  \nonumber
    && ~~~~~~~~\delta \omega_{\alpha}(z, \omega_{\alpha})  =
    \frac{\mu_0}{4 \pi \hbar} 
    \, \frac{1}{(2 z)^3} \, 
    \\ \nonumber
    && \times
    \sum_{\beta }
    \bigg \{ \, ( |\mu_x^{\alpha \beta}|^2  +  |\mu_y^{\alpha \beta}|^2  ) \, [ \, f_{\|}(k_{\beta \alpha} z) - \tilde{i}_{\|}(k_{\beta \alpha} z) \, ]  ~~~~~~~~
    \\ \label{Ee_exact}
    && ~~~~~~~~\;
    \, +  
    |\mu_z^{\alpha \beta}|^2  \, [ \, f_{\perp}(k_{\beta \alpha} z) - \tilde{i}_{\perp}(k_{\beta \alpha} z) \, ]  
    ~ \bigg \}  \, ,
\eea
   \noi
   where we have defined the dimensionless functions 
\bea   \nonumber
  f_{\|}(x) &=& - \, (2x )^2 \, ( \cos( 2x )  +  \frac{1}{2})
                \\ \nonumber
                & + & \, 2x\sin( 2x )   \, + \,   \cos( 2x  )   -  1 
		\\
		& +& \, ( \, 2x \, + \, 1   \, ) \, e^{- 2 x } ~ ,
\eea
and 
\bea
   f_{\perp}(x) &=&   2 \, f_{\|}(x)    -  2 \,   (2x)^2   \cos( \, 2x \, ) ~ . 
\eea
\\
    Here $\tilde{i}_{\rho}(k_{\beta \alpha} z)$ (using $\rho=\|,\perp$) is given by Eqs. (\ref{i_para}) and (\ref{i_perp}).
    
For short distances, i.e. $k_{\beta \alpha} z \ll 1$ (non-retarded limit), $f_{\perp}(k_{\beta \alpha} z) \approx   2 \, f_{\|}(k_{\beta \alpha} z) \approx 2$
    in which case \eq{Ee_exact} is reduced to the same result as for the state $| \beta \rangle$, i.e. \eq{omega_g_64}. Since for a two-level system $|\mu^{\alpha\beta}_a|=|\mu^{\beta\alpha}_a|$
    this agrees with the intuitive fact that the CP force for the excited ($| \alpha \rangle$) and the ground ($| \beta \rangle$) state
    should be the same in the classical limit, i.e. for $k_{\beta \alpha} z \ll 1$. In Ref.\cite{barton_74} Barton has realized the same situation for the energy shift in the two-level approximation for electric dipole transitions, i.e. the corresponding attractive short distance CP force is the same in the ground state and the excited state.

    In the long distance retarded limit, i.e. $k_{\beta \alpha}z \gg1$,  we realize that 
\bea
    f_{\|}(k_{\beta \alpha} z) \simeq - \, (2 \, k_{\beta \alpha} z )^2 \, \left( \cos( 2 \, k_{\beta \alpha} z )  +  \frac{1}{2} \right) \, ,
\eea
 and 
\bea
    f_{\perp}(k_{\beta \alpha} z) \simeq - \, (2 k_{\beta \alpha} z )^2 \, \bigg( 4 \, \cos( 2 \, k_{\beta \alpha} z )  +  1 \, \bigg) \, , 
\eea
    in which case \eq{Ee_exact} is reduced to 
\bea  \nonumber
    && \delta \omega_{\alpha}(z, \omega_{\beta })  \, \simeq \, 
    - \, \frac{\mu_0}{32 \pi \, \hbar} 
    \, \frac{k_{\beta \alpha}^2}{z} \, 
    \\ \nonumber
    && \times
    \sum_{\beta }
    \bigg \{ \, ( |\mu_x^{\alpha \beta}|^2  +  |\mu_y^{\alpha \beta}|^2  ) \, ( \, \cos( 2  k_{\beta \alpha} z )  +  \frac{1}{2} \, )  ~~~~~~~~
    \\ \label{Ee_large}
    && ~~~~~~~~~~
    \, +  
    |\mu_z^{\alpha \beta}|^2  \, ( \, 4 \, \cos( 2  k_{\beta \alpha} z )  +  1 \, )  
    ~ \bigg \}  \, .
\eea


\end{document}